\documentclass[aps,prb,twocolumn,groupedaddress,floatfix,showpacs,endfloats]{revtex4-1}
\usepackage{graphicx}
\usepackage{amsmath}
\usepackage{subcaption}

\usepackage{color}
\newcommand{\parl}{\parallel}
\begin{document}

%\preprint{}

\title{Nonlinear Optical Properties  and Self-Kerr Effect of Rydberg Excitons in Cu$_2$O}


\author{Sylwia Zieli\'{n}ska-Raczy\'{n}ska}
%\thanks{}
\author{Gerard Czajkowski}
\author{Karol Karpi\'{n}ski}
\author{David Ziemkiewicz}
\email{david.ziemkiewicz@utp.edu.pl}

 \affiliation{Institute of
Mathematics and Physics, UTP University of Science and Technology,
\\ Al. Prof. S. Kaliskiego 7, PL 85-789 Bydgoszcz, Poland}


\date{\today}

\begin{abstract}  We show how to compute the nonlinear optical  functions (absorption, reflection, and transmission) for a medium with {Rydberg
excitons}, including the effect of the coherence between
the electron-hole pair and the electromagnetic  field. Using the
Real Density Matrix Approach the analytical expressions for
nonlinear optical functions are obtained and numerical
calculations for Cu$_2$0 crystal are performed.  We report a good
agreement with recently published experimental data. Propagation
of the electromagnetic waves in Rydberg excitons media with
nonliner effect is also discussed and the possibility of obtainig
self-phase modulation due to Kerr nonlinearity is
investigated.\end{abstract}

\pacs{78.20.−e, 71.35.Cc, 71.36.+c}
%\keywords{}

\maketitle

\section{Introduction}

Recently, a lot of attention has been drawn back to the subject of
excitons in bulk crystals due to an experimental observation of
the so-called yellow exciton series in Cu$_2$O up to a large
principal quantum number of \emph{n} = 25. \cite{Kazimierczuk}
Such excitons in copper oxide, in analogy to atomic physics, have
been named Rydberg excitons (RE). By virtue of their special
properties Rydberg excitons have become widely explored in solid
and optical physics. These objects whose size scales as the square
of the principal quantum number \emph{n}, are ideally suited for
fundamental quantum interrogations, as well as detailed classical
analysis. Due to their exaggerated properties, including long
lifetimes, large electric dipole moments, and strong
exciton-exciton interactions controlled by so-called Rydberg
blockade, REs could have promising applications in, among many
areas, quantum calculating and quantum information.

 Several
theoretical approaches to calculate optical properties of REs
 have been presented (see, for example, Ref.\cite{FK} for
a review). As was observed just in the first experiment,
\cite{Kazimierczuk} the optical line shapes of RE are very
sensitive to  laser power of the exciting electromagnetic wave,
which could be connected with nonlinear effects. However, almost
all efforts in the area of RE have been mainly devoted to the
linear optical properties of Rydberg excitons. The nonlinear
phenomena with RE were first discussed by Walther \textit{et al}.
regarding coupling between strong interaction of RE and optical
photons in semiconductor microcavity. \cite{Walther} Also the
recent experiments by Heckt\"{o}tter \emph{et
al}.\cite{Hecktoetter_2018} has touched the problem of nonlinear
properties of RE in the presence of electron and hole plasma and
paved the way
 to extend the  discussion of the
optical properties of RE  to include the nonlinear effects.

 In this paper we
discuss the role of the  exciting light
intensity on the RE spectra, considering the impact of
simultaneous inter-band and intra-band excitations. As in our
previous paper, \cite{Zielinska.PRB} we will use the method based
on the Real Density Matrix Approach (RDMA). While the experiments
with REs were performed at helium temperatures there are in
principle no obstacles for observations in temperatures reaching
100 K. \cite{Stolz}$^,$\cite{Kitamura}
  By taking into account density matrices
  for electrons and holes we are able to include additional  effects  regarding temperature, which influences
   the gap energy as well as intrinsic damping parameter. \cite{Stolz}$^,$\cite{Modulator} We derive
  expressions for the  third order susceptibility
   $\chi^{(3)}$ which enable us to obtain formula for the nonlinear optical functions. The calculation are performed  to a
   Cu$_2$O crystal for which
   the nonlinear optical functions in the case of quasi-stationary excitation are analytically calculated.

Rydberg excitons in coprous oxide might be of
great potential for future application in photonic quantum
information processing, where nonlinear interaction plays the
crucial role, so we have directed our interest to light
propagation in this medium. The electromagnetic waves propagation
in nonlinear regime discussed in the present paper, is connected
with
 Kerr effect, which manifests itself as  a self-induced phase of a pulse of light as it travels through
  the medium. Self-phase modulation (SPM) is a nonlinear optical effect of light-matter interaction;  it is induced by a
  varying refractive index of the medium. This  produces a phase shift in the pulse, leading to a
   change of the pulse's frequency spectrum.   It might be interesting
   to examine the intensity dependence of the index of refraction in REs media.
   Such approach might be the first step to develop the new branch of investigations and applications in media with RE; the phenomenon of self-focusing is the promising example. \cite{self-foc}
SPM steered on demand by light intensity can also find interesting applications in optoelectronic devices working with low-light intensity,
 such as all-optical switching and logic gates.
 In quantum information context two-photon self-Kerr nonlinearities may be used in quantum computing. \cite{Joshua}

The paper is organized as follows. In
     Sec.~\ref{density.matrix} we recall the basic equations of
     the RDMA and formulate the equations for the case when the
     inter-band and intra-band electronic transitions are accounted for. In Sec.~\ref{iteration}
     we describe an iteration procedure, which will be applied to
     solve a system of coupled integro-differential equations. The second iteration step,
     from which the nonlinear optical functions will be calculated, is given in
     Sec.~\ref{iteration2nd}. The formulas, derived in this
     Section, are than applied in Sec~\ref{calculations}, which is devoted to presentation and discussion of the
     nonlinear optical functions
 for a Cu$_2$O crystal. Self-phase modulation in such a system regarding intensity and transmission dependence is  discussed in Sec. \ref{Kerr}. Finally, in Sec.~\ref{conclusions} we
     draw conclusions of nonlinear studies presented in this paper.

\section{Excitonic Nonlinearities in Rydberg Exciton Media: Density
Matrix Formulation}\label{density.matrix}

 In what follows we adapt the real density matrix approach to the case
of semiconductors under high excitation, and show how to calculate
the nonlinear optical functions.

 We discuss the nonlinear response of a semiconductorslab
to an  electromagnetic wave characterized by the electric field
vector,
\begin{equation}
{\bf E}=\textbf{E}_{i0}\exp( i\textbf{k}_0{\textbf{R}}-i\omega t),
\quad k_0=\omega/c,
\end{equation}
\textbf{R} being the excitonic center-of-mass coordinate (see Eq.
(\ref{com}) below).

 In the
RDMA approach the bulk nonlinear response will be described by a
closed set of differential equations ("constitutive equations"):
one for the coherent amplitude $Y({\bf r}_1, {\bf r}_2)$
representing the exciton density related to the interband
transition, one for the density matrix for electrons $C({\bf r}_1,
{\bf r}_2)$ (assuming a non-degenerate conduction band), and one
for the density matrix for the holes in the valence band, $ D({\bf
r}_1, {\bf r}_2)$. Below we will use the notation
\begin{equation}
 Y({\bf r}_1, {\bf r}_2)=Y_{12},\quad\hbox{etc}.
\end{equation}
\noindent The constitutive equations have the form: the interband
equation
\begin{eqnarray}\label{interband1}
 & &{
i}\hbar\partial_tY_{12}-H_{eh}Y_{12}=-{\bf M}{\bf E}({\bf
R}_{12})\nonumber
\\
& &+{\bf E}_1{\bf M}_{0}C_{12}+{\bf E}_2{\bf M}_{0}D_{12}+{
i}\hbar\left(\frac{\partial Y_{12}}{\partial t}\right)_{{\rm
irrev}},
\end{eqnarray}
conduction band equation\begin{eqnarray}
& &\label{conduction1}{i}\hbar\partial_tC_{12}+H_{ee}C_{12}={\bf M}_{0}({\bf E}_1 Y_{12}-{\bf E}_2Y^*_{21})\nonumber\\
& &+{i}\hbar\left(\frac{\partial C_{12}}{\partial t}\right)_{{\rm
irrev}},\end{eqnarray} valence band equation
\begin{eqnarray} &
&\label{valence1}{i}\hbar\partial_tD_{21}-H_{hh}D_{21}={\bf
M}_{0}({\bf E}_2 Y_{12}-{\bf E}_1Y^*_{21})\nonumber\\
&&+ {i}\hbar\left(\frac{\partial D_{21}}{\partial t}\right)_{{\rm
irrev}},
\end{eqnarray}
\noindent where the operator $H_{eh}$ is the  effective mass
Hamiltonian
\begin{equation}\label{hamiltonianeh}
H_{eh}=E_g-\frac{\hbar^2}{2M_{tot}}\partial_Z^2-\frac{\hbar^2}{2M_{tot}}\hbox{\boldmath$\nabla$}_{R_\parl}^2-\frac{\hbar^2}{2\mu}\partial_z^2-\frac{\hbar^2}{2\mu}\hbox{\boldmath$\nabla$}_\rho^2+V_{eh},
\end{equation}
\noindent with the separation of the center-of-mass coordinate
${\bf R}_\parl$ from the relative coordinate
$\hbox{\boldmath$\rho$}$ on the plane $x-y$,
\begin{eqnarray}
&
&H_{ee}=-\frac{\hbar^2}{2m_e}(\hbox{\boldmath$\nabla$}_1^2-\hbox{\boldmath$\nabla$}_2^2),\nonumber
\\
&
&H_{hh}=-\frac{\hbar^2}{2m_{h}}(\hbox{\boldmath$\nabla$}_1^2-\hbox{\boldmath$\nabla$}_2^2),
\end{eqnarray}
\noindent and ${\bf E}_{12}$ means that the wave electric field in
the medium is taken in a middle point between ${\bf r}_1$ and
${\bf r}_2$: we take them at the center-of-mass
\begin{equation}\label{com}
{\bf R}={\bf R}_{12}=\frac{m_h{\bf r}_1+m_e{\bf r}_2}{m_h+m_e}.
\end{equation}
\noindent In the above formulas $m_e, m_h$ are the electron and
the hole effective masses (more generally, the effective mass
tensors), $M_{tot}$ is the total exciton mass, and $\mu$ the
 reduced mass of electron-hole pair. The smeared-out transition
dipole density ${\bf M}({\bf r})$ is related to the bilocality of
the amplitude $Y$ and describes the quantum coherence between the
macroscopic electromagnetic field and the inter-band transitions.
\cite{StB87} The resulting
coherent amplitude $Y_{12}$ determines the excitonic part of the
polarization of the medium
\begin{eqnarray}\label{polarization1}
&&{\bf P}({\bf R},t)=2\int{\rm d}^3r\,\textbf{M}^*({\bf
r})\hbox{Re}~Y({\bf
R},{\bf r},t)\nonumber\\
&&=\int{\rm d}^3r\textbf{M}^*({\bf r})[Y({\bf R},{\bf
r},t)+\hbox{c.c}],
\end{eqnarray}
\noindent where ${\bf r}={\bf r}_1-{\bf r}_2$ is the electron-hole
relative coordinate.

The linear optical properties are obtained by solving the
interband equation (\ref{interband1}), supplemented by the
corresponding Maxwell equation, where the polarization
(\ref{polarization1}) acts as a source. For computing the
nonlinear optical properties we use the entire set of constitutive
equations (\ref{interband1})-(\ref{valence1}). At the moment a
general solution of the equations seems to be inaccessible. Only
in special situations a solution can be found. For example, if one
assumes that the matrices $Y,C$ and $D$ can be expanded in powers
of the electric field ${\bf E}$, an iteration scheme can be used.

The relevant expansion of the polarization in powers of the field
has the form
\begin{equation}
P({\bf k},\omega)=\epsilon_0\left[ \chi^{(1)}E({\bf k},\omega)+
\chi^{(3)}(\omega,-\omega,\omega)\vert E({\bf
k},\omega)\vert^2+\ldots\right],
\end{equation}
\noindent where $\chi^{(1)}$ is the linear and
$\chi^{(3)}$ is a nonlinear susceptibility.

\section{The iteration procedure} \label{iteration}
\noindent Following the calculation scheme proposed in Refs. \cite{StB87,FrankStahl,RivistaGC}, we calculate the susceptibility
$\chi^{(3)}$ iteratively from the dynamic equations
(\ref{interband1})-(\ref{valence1}). The first step in the
iteration consists of solving the equation (\ref{interband1}),
which we take in the form
\begin{equation}\label{Ylin}
{i}\hbar\partial_tY^{(1)}_{12H}-H_{eh}Y^{(1)}_{12}=-{\bf M}{\bf
E}+{i}\hbar\left(\frac{\partial Y_{12}^{(1)}}{\partial
t}\right)_{{\rm irrev}}.
\end{equation}
\noindent For the irreversible part we assume, as usually, a
relaxation time approximation
\begin{equation}
\left(\frac{\partial Y_{12}^{(1)}}{\partial t}\right)_{{\rm
irrev}}=-\frac{1}{T_{2}} Y_{12} = \frac{-\Gamma}{\hbar}Y_{12}.
\end{equation}
where $\mit\Gamma$ is a dissipation constant and $T_{2}=\hbar/{\mit\Gamma}$. In the discussion of nonlinear effects we take also into
account the non-resonant parts of the  amplitude $Y$, and consider
the electric field ${\bf E}$ in the medium in the form

\begin{equation}
{\bf E}_{12}={\bf E}({\bf R},t)+{\bf E}^*({\bf R},t)= {\bf
E}_0e^{{ i}({\bf kR}-\omega t)} + {\bf E}_0e^{-{ i}({\bf
kR}-\omega t)}.
\end{equation}
\noindent Therefore the equations (\ref{Ylin}) generate two
equations: one for an amplitude $Y_{-}^{(1)}\,\propto \exp(-{
i}\omega t)$, and the second for the non resonant part
$Y^{(1)}_{+}\,\propto \exp({ i}\omega t)$,

\begin{eqnarray}
& &{i}\hbar\left( i\omega+\frac{1}{T_{2}}\right)Y^{(1)}_{12+}-
H_{eh}Y^{(1)}_{12+}=-{\bf M}{\bf E}^*({\bf R},t),\nonumber
\\
&&\\ & &{i}\hbar\left(-{
i}\omega+\frac{1}{T_{2}}\right)Y^{(1)}_{12-}-
H_{eh}Y^{(1)}_{12-}=-{\bf M}{\bf E}({\bf R},t).\nonumber
\end{eqnarray}
We will consider only one component of ${\bf E}$ and
${\bf M}$. As in our previous papers\cite{Zielinska.PRB,Zielinska.PRB.2016.c} we look for a
solution in terms of eigenfunctions of the Hamiltonian $H_{eh}$,
which we use in the form $\varphi_{n\ell m}=R_{n\ell}(r)Y_{\ell m}(\theta,\phi)$, where $R_{n\ell}$ are the hydrogen radial functions, and $E_{n}$ the corresponding eigenvalues.
 So we obtain
\begin{eqnarray}\label{yplusminus}
&&Y_{12-}=E({\bf R},t)\sum_n\frac{c_{n\ell m}\varphi_{n\ell m}({\bf r})}{\hbar(\Omega_{n\ell m}-\omega-{ i}/T_{2n})},\nonumber\\
&&Y_{12+}=E^*({\bf R},t)\sum_n\frac{c_{n\ell m}\varphi_{n\ell
m}({\bf r})}{\hbar(\Omega_{n\ell m}+\omega-{i}/T_{2n})},
\end{eqnarray}
where
\begin{eqnarray}\label{basic}
&&c_{n\ell m}=\int{\rm d}^3r M({\bf r})\varphi_{n\ell m}({\bf r}),\nonumber\\
&&\hbar\Omega_{n\ell m}=E_{glm}+E_{n}+\frac{\hbar^2}{2M_{tot}}k_z^2,\\
&&r=\sqrt{x^2+y^2+z^2},\nonumber
\end{eqnarray}
and, in the case of Cu$_2$O,  $E_{g\ell m}$ are the gap energies
appropriate for $p$ and $f$ excitons. For the sake of simplicity,
we consider only the $p$ excitons contribution.  The solutions
$Y_{12\pm}^{(1)}$ determine the linear susceptibility
\begin{eqnarray}\label{chilin}
& &\chi^{(1)}\left(\omega,k_z^{(1)}\right)=\frac{1}{\epsilon_0E_0}\int {\rm d}^3{\bf r}\left[Y_{12-}^{(1)}+Y_{12+}^{(1)*}\right]M^*({\bf r})\nonumber\\
&&=\frac{1}{\epsilon_0\hbar}\sum_{n\ell
m}\frac{b_{n1}\Omega_{n\ell m}}{\Omega_{n\ell m}^2-(\omega+{
i}/T_{2n})^2},
\end{eqnarray}
\noindent with the coefficients\cite{Zielinska.PRB}
\begin{eqnarray}\label{bn1}
&&b_{n1}=2\vert c_{n10}\vert^2=\nonumber\\
&&=\frac{8\pi}{3}\left(\int\limits_0^\infty r^2 {\rm d}r
M(r)R_{n1}(r)\right)^2.
\end{eqnarray}

 The so obtained susceptibility
defines the linear dispersion rule for the polariton modes
\begin{equation}
\frac{c^2\left(k_z^{(1)}\right)^2}{\omega^2}=\epsilon_b+\chi^{(1)}\left(\omega,k_z^{(1)}\right).
\end{equation}
The coefficients $b_{n1}$ can be expressed in terms of the band
parameters and, for energies below the gap, one obtains
\begin{eqnarray}\label{polariton1}
\frac{k_z^{(1)2}}{k_0^2}-\epsilon_b=\hspace{15em}\\
=\epsilon_b\sum\limits_{n=2}^N\frac{f_{n1}\Delta_{LT}/R^*}{\left(E_{Tn}-E-{
i}{\mit\Gamma_n}\right)/R^*
+(\mu/M_{tot})(k_z^{(1)}a^*)^2}\nonumber
\end{eqnarray}
where $k_0=\omega/c$, ${\mit\Gamma}_n=\hbar/T_{2n}$,  $E_{Tn}$ are
energies of the exciton resonances, and the oscillator strengths are given by\cite{Zielinska.PRB}
\begin{equation}\label{oscillatorforces}
f_{n1}=\frac{32(n^2-1)}{3n^5}\left[\frac{n(r_0+2a^*)}
{2(r_0+na^*)}\right]^6.
\end{equation}
For the wave propagating in the medium, we choose the appropriate
component of the dipole density. For $p$ excitons ($\ell=1$) the
$z$-component of the dipole density vector, which will be used
below, has the form \cite{Zielinska.PRB}

\begin{eqnarray}\label{Mz1}
&&M_z({\bf r})=M_{10}\frac{r+r_0}{2r^2r_0^2}
\sqrt{\frac{4\pi}{3}}Y_{10}e^{-r/r_0},
\end{eqnarray}
with the coherence radius $r_0$  \cite{StB87,Zielinska.PRB}
\begin{equation}\label{r0}
r_{0}^{-1}=\sqrt{\frac{2\mu_\parl}{\hbar^2}E_g}.
\end{equation}

\section{Iteration procedure: second step}\label{iteration2nd}

Let us  consider a wave linearly polarized in the $z$ direction.
 Then $Y_{\pm}^{(1)}$ (\ref{yplusminus}) are inserted into the source terms of the
conduction-band and valence band equation (\ref{conduction1} -
\ref{valence1}).  Solving for stationary solutions, we have
\begin{eqnarray}
&
&J_{C}=M_{10}\left(E_1Y_{12}^{(1)}-E_2Y_{21}^{(1)*}\right)\nonumber
\\
& &=\frac{2{i}M_{10}{E}_0^2}{\hbar}\left[\exp\left({
i}k_z^{(1)}z\frac{m_{e}}{M_{tot}}\right)\,\hbox{Im}\,g(-\omega,
{\bf
r})\right.\\
&&\left.+ \exp\left(-{
i}k_z^{(1)}z\frac{m_{e}}{M_{tot}}\right)\,\hbox{Im}\,g(\omega,
{\bf r})\right],\nonumber
\end{eqnarray}
\noindent where\begin{equation} g(\pm\omega, {\bf
r})=\sum_{j}\frac{c_{j10}\varphi_{j10}({\bf
r})}{\Omega_{j10}\mp\omega-{i}/T_{2j1}}.
\end{equation}
\noindent For the source terms of the valence band equations
(\ref{valence1}) we obtain
\begin{eqnarray}
& &J_{V}=M_{01}(E_2Y_{12}^{(1)}-E_1Y_{21}^{(1)*})\nonumber\\ &
&=\frac{2{i}M_{10}{E}_0^2}{\hbar}\left[\exp\left({
i}k_z^{(1)}z\frac{m_{h}}{M_{tot}}\right)\,\hbox{Im}\,g(\omega,
{\bf
r})\right.\\
&&\left.+ \exp\left(-{
i}k_z^{(1)}z\frac{m_{h}}{M_{tot}}\right)\,\hbox{Im}\,g(-\omega,
{\bf r})\right].\nonumber
\end{eqnarray} If
irreversible terms are well defined, the equations
(\ref{valence1}) can be solved and their solutions are then used
in the saturating terms on the r.h.s. of the equations
(\ref{interband1}). Assuming relaxation time approximation the  time dependence of density matrices $C$ and $D$ is described as
\begin{eqnarray}\label{relaxation_time}
& &\left(\frac{\partial C}{\partial t}\right)_{{\rm
irrev}}\nonumber\\
&& =-\frac{1}{\tau}\left[C({\bf X},{\bf r},t)-f_{0e}({\bf
r})C({\bf X}, {\bf
r}=\textbf{r}_0,t)\right]-\frac{C(r_0)}{T_1},\nonumber
\\
&&\\
 & &\left(\frac{\partial D}{\partial t}\right)_{{\rm
irrev}}\nonumber\\
&&= -\frac{1}{\tau}\left[D({\bf X},{\bf r},t)-f_{0h}({\bf
r})D({\bf X}, {\bf
r}=\textbf{r}_0,t)\right]-\frac{D(r_0)}{T_1},\nonumber
\end{eqnarray}
\noindent where
\begin{equation}
{\bf X}=\frac{1}{2}\left({\bf r}_e+{\bf r}_h\right),
\end{equation}
\noindent and $f_{0e}, f_{0h}$ are normalized Boltzmann
distributions for electrons and holes, respectively and  $\tau$ denotes the relaxation time. The  relaxation $T_1$ stands for interband
recombination\cite{FrankStahl} and $f_{0e}$ is defined as
\begin{eqnarray*}
&&f_{0e}({\bf r})=\int {\rm d}^3q f_{0e}({\bf q})e^{-{i}{\bf qr}}
=\exp\left[-\frac{m_e{ k_B{\mathcal T}}}{2\hbar^2}r^2\right]
\end{eqnarray*}
\noindent with
\begin{equation}
f_{0e}({\bf q})=\left(\frac{\hbar^3}{2\pi{ k_B {\mathcal
T}}}\right)^{3/2}\frac{1}{m_e^{3/2}}\exp\left(-\frac{\hbar^2q^2}{2m_e{
k_B {\mathcal T}}}\right),
\end{equation}
where $\mathcal T$ is the temperature and $k_B$ is the Boltzmann
constant. Similarly, for the hole equilibrium distribution, we
have
\begin{equation}
f_{0h}({\bf r})=\exp\left(-\frac{m_{h}{ k_B {\mathcal
T}}}{2\hbar^2}r^2\right).
\end{equation}
\noindent Matrices $C$ and $D$ are related to charge densities
\begin{equation}
\rho_e=-eC({\bf r},{\bf r}),\qquad \rho_h=e D({\bf r},{\bf r}),
\end{equation}
which are conserved quantities. We therefore have assumed
that they relax to an equilibrium normalized to the actual number
of carriers. It should be noticed that matrices $C$
and $D$ are temperature-dependent and they give an
additional contribution for interpretation of temperature
variations of excitonic optical spectra. However, the temperature dependence of relaxation constants $\Gamma_n$ remains a dominant mechanism influencing the spectra. Further, we will assume
that our medium is excited homogeneously in ${\bf X}$- space. For
$p$ excitons the matrices $C$ and $D$ relax to their values at
$r=r_0$. In Cu$_2$O, the dipole
density can be approximated by\cite{StB87} $\textbf{M}(\textbf{r})\propto
\textbf{r}\delta(r-r_0)$, which leads to the
following expressions for the matrices $C,~D$
\begin{eqnarray}
&&C({\bf r})=-\frac{i}{\hbar}\left[\tau J_C({\bf r})-\tau
J_C(r_0)+T_1f_{0e}({\bf r})J_C(r_0)\right],\nonumber\\
&&\\ & &D({\bf r})=-\frac{i}{\hbar}\left[\tau J_{V}({\bf r})-\tau
J_{V}(r_0)+T_{1}f_{0hH}({\bf r})J_{V}(r_0)\right].\nonumber
\end{eqnarray}
With above expressions the equation for the third order coherent amplitude
$Y^{(3)}_{12}$ takes the form
\begin{eqnarray}\label{jotplus}
& &\hbar\left(\omega+\frac{\rm
i}{T_{2}}\right)Y^{(3)}_{12-}-H_{eh}
Y^{(3)}_{12-}\nonumber\\
&&=M_{10}({E}_1C_{12}+{E}_2D_{21})={E}({\bf
R},t)\tilde{J}_{-},\nonumber
\\
& &\hbar\left(-\omega+\frac{\rm
i}{T_{2}}\right)Y^{(3)}_{12+}-H_{eh}
Y^{(3)}_{12+}\nonumber\\
&&=M_{10}({E}^*_1C_{12}+{E}^*_2D_{21})={E}^*({\bf
R},t)\tilde{J}_{+},
\end{eqnarray}
\noindent where
\begin{eqnarray}
& &\tilde{J}_{-}=-\frac{i}{\hbar}M_{10}\biggl\{\tau\biggl[J_C({\bf r})e^{{i}k_zzm_{e}/M_{z}}\nonumber\\
&&+J_{V}({\bf r})e^{-{ i}k_zzm_{h}/M_{tot}}\biggr]\\
&&-\tau J_C(r_0)e^{{ i}k_zzm_{e}/M_{z}}-\tau J_{V}(r_0)e^{-{ i}k_z^{(1)}zm_{h}/M_{tot}}\nonumber\\
& &+T_{1}J_C(r_0)f_{0e}({\bf r})e^{{
i}k_z^{(1)}zm_{e}/M_{tot}}\nonumber\\
&&+T_{1}J_{V}(r_0)f_{0h}({\bf r})e^{-{
i}k_z^{(1)}zm_{h}/M_{tot}}\biggr\},\nonumber
\end{eqnarray}
\begin{eqnarray}
& &\tilde{J}_{+}=-\frac{ i}{\hbar}M_{10}\biggl\{\tau\biggl[J_C({\bf r})e^{-{i}k_z^{(1)}zm_{e}/M_{tot}}\nonumber\\
&&+J_{V}({\bf r})e^{{ i}k_z^{(1)}zm_{h}/M_{tot}}\biggr]\\
&&-\tau J_C(r_0)e^{-{\rm i}k_z^{(1)}zm_{e}/M_{tot}}-\tau J_{V}(r_0)e^{{i}k_z^{(1)}zm_{h}/M_{tot}}\nonumber\\
& &+T_{1}J_C(r_0)f_{0e}({\bf r})e^{-{\rm
i}k_z^{(1)}zm_{e}/M_{tot}}\nonumber\\
&&+T_{1}J_{V}(r_0)f_{0h}({\bf r})e^{
ik_z^{(1)}zm_{h}/M_{tot}}\biggr\}.\nonumber
\end{eqnarray}
 From $Y^{(3)}$ one finds the third order polarization
according to

\begin{eqnarray}
&&P^{(3)}({\bf R})=2\int {\rm d}^3{r} \hbox{Re}\, M({\bf r})
Y^{(3)}({\bf R},{\bf r})\nonumber
\\&&=\int {\rm d}^3{r}\,M({\bf r})\left(Y^{(3)}_{12-}+Y^{(3*)}_{12+}\right).
\end{eqnarray}
\noindent The fact that the source terms $\tilde{J}_{\pm}$ contain
terms proportional either to the relaxation time $\tau$ or to the
interband recombination  time $T_1$ allows a further
approximation. For most semiconductors
$T_1\gg\,\tau$, so the terms proportional to $\tau$ can be
neglected. As in the case of linear amplitudes $Y^{(1)}$, we
expand the nonlinear amplitudes in terms of the eigenfunctions
$\varphi_{n\ell m}({\bf r})$ obtaining
\begin{eqnarray}\label{chi31}
& &\chi^{(3)}(\omega,k_z^{(1)})\nonumber
\\&&=-
\frac{M_{10}^2T_{1}}{\epsilon_0\hbar^3}\sum_{n}\frac{ c_{n1
0}\Omega_{n10}}{\Omega_{n10}^2-(\omega+{
i}T_{2}^{-1})^2}\nonumber\\
&&\times
\Biggl\{\left[\hbox{Im}\,g(\omega,r_0)+\hbox{Im}\,g(-\omega,r_0)\right]\\&
&\times\, \left\langle\varphi_{n10}\vert e^{{
i}k_z^{(1)}z\frac{m_{e}}{M_{tot}}}f_{0e}({\bf r})+e^{-{
i}k_z^{(1)}z\frac{m_{h}}{M_{tot}}}f_{0h}({\bf
r})\right\rangle\Biggr\},\nonumber
\end{eqnarray}
\noindent where
\begin{equation}
\langle\varphi_{n10}\vert f\rangle=\int {\rm d}^3{r}\,
\varphi_{n10}({\bf r})f({\bf r}).
\end{equation}
Assuming further that $\hbar\omega$ is just below the band edge,
with regard to the relation (\ref{bn1}), and
\begin{eqnarray}&&\Delta_{LT}=\frac{\pi}{\epsilon_0\epsilon_b
a^{*3}}M_{10}^2\left(\frac{a^*}{r_0}\right)^4\left(\frac{2r_0}{r_0+2a^*}\right)^6\nonumber\\
&=&R^*\cdot
2\frac{2\mu}{\hbar^2}\frac{M_{01}^2}{\pi\epsilon_0\epsilon_ba^*}f(r_0,a^*),\end{eqnarray}
 one obtains
\begin{eqnarray}\label{chifinal}
&&\chi^{(3)}=-\frac{8\pi\epsilon_0(\epsilon_b\Delta_{LT})^2a^{*3}}{g_2(r_0)}\\
&&\times\sum\limits_{jn}\left(\frac{{\mit\Gamma}_j}{{\mit\Gamma}_{01}}\right)
\frac{\varphi_{j10}(\rho_0)\sqrt{f_{j1}f_{n1}}E_{Tn1}(A_{n10}+B_{n10})}{[(E_{Tj1}-E)^2+{\mit\Gamma_j}^2][E_{Tn1}^2-(E+i{\mit\Gamma}_n)^2]},\nonumber
\end{eqnarray}
\noindent with coefficients
\begin{eqnarray}\label{aenha}
&& A_{n10}=
\langle\varphi_{n10}\vert e^{{ i}k_{ez}z}f_{0e}({\bf r})\rangle,\nonumber\\
&& B_{n10}=\langle\varphi_{n10}\vert e^{-{ i}k_{hz}z}f_{0h}({\bf
r}\rangle,
\end{eqnarray}
where

\begin{eqnarray}\label{keh}
&&k_{ez}=k_{jz}^{(1)}\frac{m_e}{M_{tot}},\quad
k_{hz}=k_{jz}^{(1)}\frac{m_{h}}{M_{tot}},\nonumber\\
 &&{\mit\Gamma}_j=\frac{\hbar}{T_{2j}},\qquad
{{\mit\Gamma}_{01}}=\frac{\hbar}{T_1},\nonumber\\
&&\varphi_{j10}(\rho_0)=\sqrt{\frac{3}{4\pi}}R_{j1}(\rho_0),\quad \rho_0=\frac{r_0}{a^*},\\
&&E_{Tn1}=\hbar\Omega_n(k_z=0),\qquad
E=\hbar\omega.\nonumber\end{eqnarray}
\section{Nonlinear optical functions}\label{calculations}
Basing on the nonlinear susceptibility $\chi^{(3)}$, one obtains the total index of refraction
\begin{eqnarray}\label{indexn3}
\left[n^{(3)}\right]^2=\epsilon_b\biggl[1+\frac{\chi^{(1)}}{\epsilon_b}
+\vert E_{prop}\vert^2\frac{\chi^{(3)}}{\epsilon_b}\biggr],
\end{eqnarray}
 where $E_{prop}$ is the amplitude of the wave propagating in the crystal. It is obtained from the equation
\begin{eqnarray}\label{intensity}
&&\vert
E_{prop}\vert^2=2\left|\frac{2}{1+\sqrt{\epsilon_b}}\right|^2\zeta
P,
\end{eqnarray}
where $P$ is the laser power, and $\zeta\approx 377\,\Omega$
is the impedance of free space.

 Regarding the experiment by Heckt\"{o}tter \emph{et al}.\cite{Hecktoetter_2018} carried out at the temperature 1.35 K, we performed the calculations for a Cu$_2$O crystal of thickness 40 $\mu$m,
 for various powers of the impinging light.
 The band parameters of cuprous oxide used in the calculations are collected in Table
 \ref{parametervalues}. We used the damping parameters
 ${\mit\Gamma}_n$ obtained by fitting of the experimental curves
 by Kazimierczuk \emph{et al}. \cite{Kazimierczuk} and Heckt\"{o}tter \emph{et
 al}.
 \cite{Hecktoetter_2018}

\begin{figure}[h]
\centering
\includegraphics[width=1\linewidth]{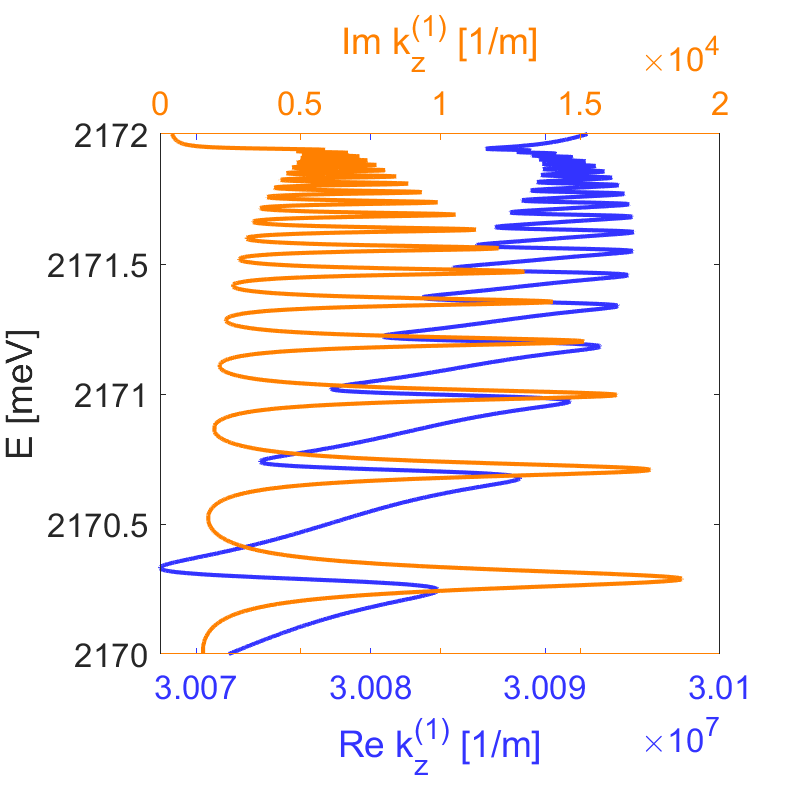}
\caption{The real and imaginary part of the wave vector
$k_z^{(1)}$ of a Cu$_2$O crystal, in the energetic region of
$n=2-20$ excitonic states, calculated by Eq. (\ref{polariton1}).}
\label{klinear}
\end{figure}

We start the iteration procedure with the linear polariton
problem, taking in the bulk dispersion
 the antiresonant terms as in the Eq.~(\ref{chilin}). We obtain the polariton wave vector $k_{z}^{(1)}$ from Eq. (\ref{polariton1}) and
the bulk polariton amplitude $E_{prop}$. The real and imaginary
part of the wave vector are presented in Fig.\ref{klinear}.
The relation has been calculated without
accounting the spatial dispersion, so we obtained $k_z^{(1)}$ as a
one-valued function of the wave energy.  The
spatial dispersion gives a set of polariton branches and distinct
polariton waves. The calculations require the valus of amplitudes
of the polariton waves. This would lead to the problem of
Additional Boundary Conditions (ABC) in the case of a large number
of polaritons (as appears in Cu$_2$O) which goes beyond the scope
of the present paper.

When the spatial dispersion effects are neglected, the amplitude
of the propagating wave results from the standard relation for the
half-space geometry (\ref{intensity}).

 In the second
step we compute the third order susceptibility and the total
excitonic bulk polarization using
 as input the linear polariton characteristics.

\begin{figure}[h]
\centering
\includegraphics[width=1\linewidth]{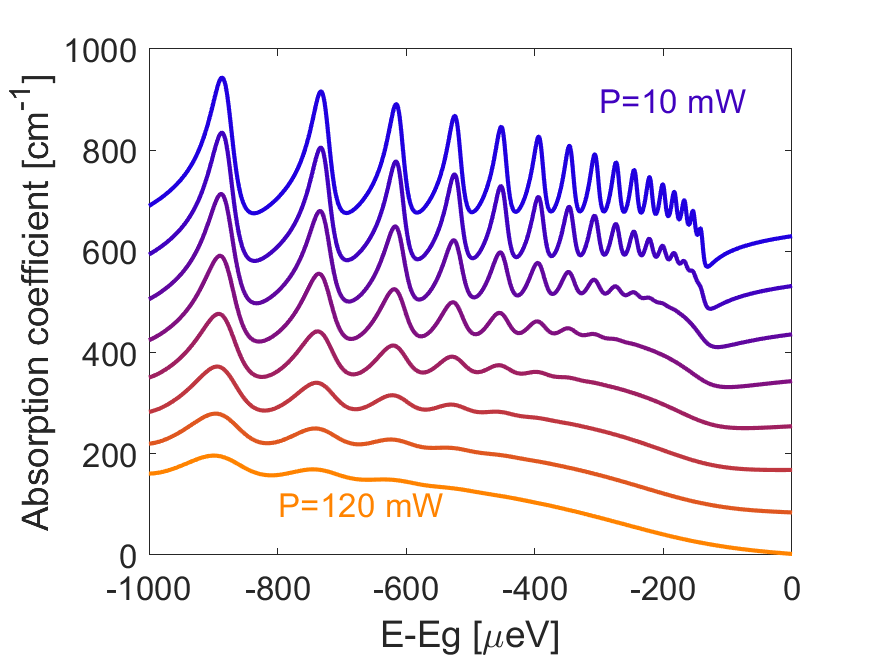}
\caption{The  nonlinear absorption coefficient of a Cu$_2$O
crystal, in the energetic region of $n=10-20$ excitonic states.}
\label{chinonlinear}
\end{figure}

 From the imaginary part one
obtains the absorption coefficient
\begin{equation}\label{alpha}
\alpha^{(3)}=2\frac{\hbar\omega}{\hbar c}\hbox{Im}\,n^{(3)}.
\end{equation}
It has been calculated for various laser intensities, in the
energetic region of the $n=10,\ldots 20$ excitonic states
(Fig.\ref{chinonlinear}). The results are in
excellent agreement with the experimental data by Hecktotter
\emph{et al}. \cite{Hecktoetter_2018}; the maxima of absorption
are vanishing for higher excitonic states, especially for larger
laser power. Additionally, the optical bleaching is noticeable, e.
g. the overall absorption is decreasing with power. An intensity
dependent shift of the positions of the excitonic resonances can
be attributed to the interplay between the real and imaginary part
of the susceptibility. Here one can notice the advantage of using
RDMA which gives simultaneously (without use of Kramers-Kronig
relations) both parts of the susceptibility.

\begin{figure}[h]
\centering
\includegraphics[width=0.82\linewidth]{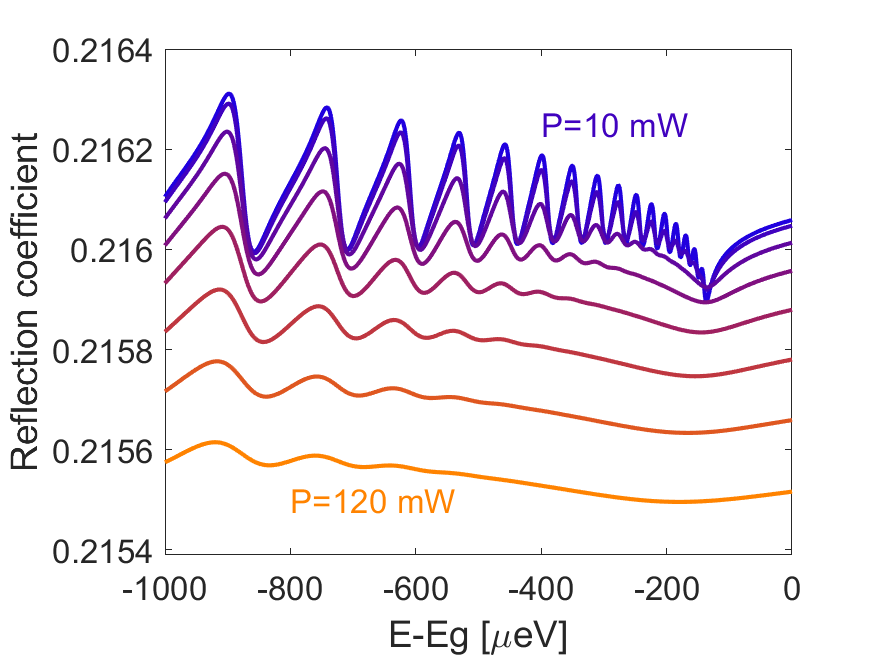}
\caption{The nonlinear reflectivity of a Cu$_2$O crystal, in the
energetic region of $n=10-20$ excitonic states.}
\label{nonlinearref}
\end{figure}

\begin{figure}[h]
\centering
\includegraphics[width=0.82\linewidth]{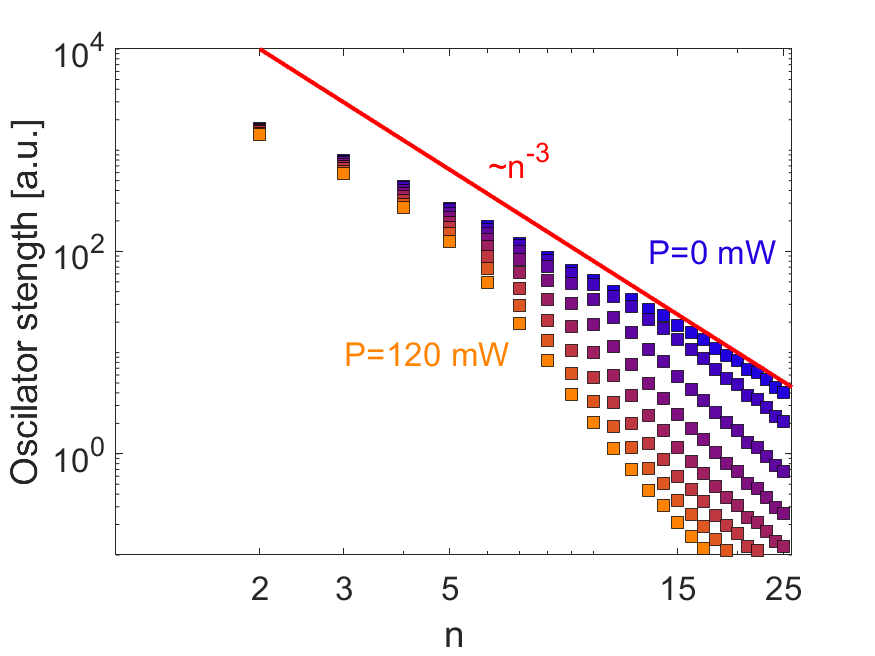}
\caption{Dependence of oscillator strength (peak area) on laser
power for different $n$ resonances.} \label{peakarea}
\end{figure}

 Finally, having the intensity dependent index of refraction (\ref{indexn3}),  we can calculate the optical
 functions (reflectivity, transmissivity, and absorption). We have chosen the reflectivity, resulting from the
 equation
 \begin{equation}\label{rnonlinear}
 R=\left|\frac{1-n^{(3)}}{1+n^{(3)}}\right|^2.
 \end{equation}
The results, for the same crystal as above, and in the same
energetic region, are shown in Fig. \ref{nonlinearref}. Both
quantities, the absorption and the reflectivity, show the
decreasing oscillator strengths, which is the main effect of the
increasing applied laser power. We have
estimated the relative oscillator strengths by integrating the peak
area for $n=1-24$ excitons,
 for various laser powers. The results are shown on the Fig. \ref{peakarea}. At low power, one obtains the well-known $n^{-3}$
 dependence reported by Kazimierczuk \emph{et al}. \cite{Kazimierczuk} evident for the Rydberg states with $n>10$.
 Our fit gives proper predictions for both $n$ number and power dependence of the oscillator strengths;
  as the power increases, the highly excited states are suppressed due to the Urbach tail. \cite{Hecktoetter_2018}
  We have approximated this effect by introducing additional relaxation term ${\mit\Gamma}' = \frac{{\mit\Gamma}_0'}{(E-Eg)^{3/2}}$
   which is added to the excitonic relaxation constants ${\mit\Gamma}_n$. The constant ${\mit\Gamma}_0'$ is obtained by fitting the data on the Fig. \ref{chinonlinear} to the experimental
  results by Heckt\"{o}tter \emph{et al}. \cite{Hecktoetter_2018}

\section{ SELF-KERR NONLINEARITY and SELF-PHASE
MODULATION}\label{Kerr}
 The self-Kerr interaction is an nonlinear self-interaction of electromagnetic wave which arises during its propagation in the medium   that produces a phase shift proportional to
the square of the field or, in quantum regime, number of photons
in the field. The dependence of the medium polarization, or
equivalently the index of refraction, on the intensity of the
field, is the base of this effect.
 The consequences of the Kerr effect is self-phase modulation (SPM); this means that a light wave in the medium  experiences a nonlinear phase change; an optical field modifies its own phase.  The
self-induced phase modulation of a pulse of light  is useful measurable parameter and the engineering of self-Kerr interaction is of great interest for processing of optical spectrum of light beams propagating through the media.

 In media which are characterized with a non-negligible  nonlinear term of
  susceptibility, the phase of a wave traversing a distance $L$ increases by $\varphi=\Delta k_z L$, and increment in phase due to the power-dependent term is equal
\begin{equation}
\varphi=\frac{\omega}{c}[n^{(3)}(P)-n^{(3)}(0)],
\end{equation}
where $n^{(3)}(P)$ is the power-dependent part of the refractive index.

\begin{figure}[h]
\begin{minipage}{\linewidth}
\centering
\includegraphics[width=0.82\linewidth]{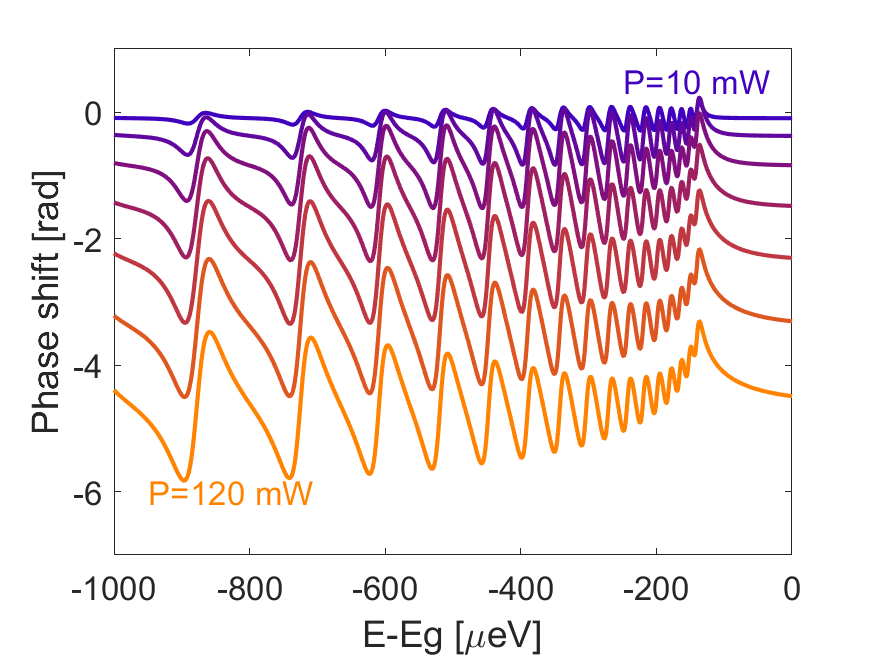}
\caption{The relative phase shift $\varphi(P)-\varphi(0)$, for
various laser powers $P$ and crystal length $L=100\;\mu\hbox{m}$.}
\label{nonlinearphi}
\end{minipage}
\end{figure}

\begin{figure}[h]
\centering
\includegraphics[width=1\linewidth]{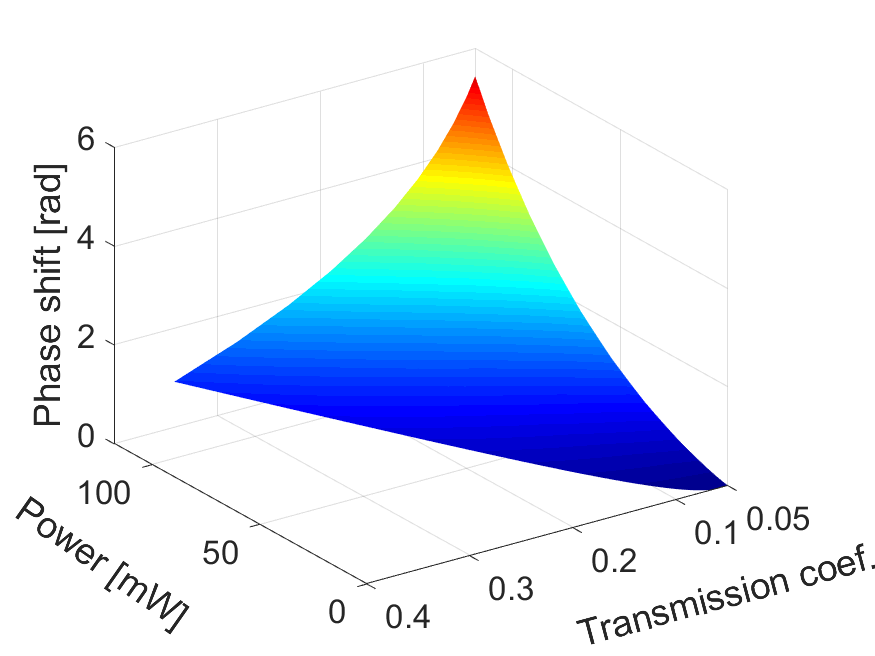}
\caption{The absolute value of phase shift as a function of
absorption coefficient and laser power, for $E = 2171\;
\hbox{meV}$.} \label{nonlinearphi2}
\end{figure}

It seems that coprous oxide with Rydberg excitons is a superior material for solide-state quantum optics,
 so we have performed numerical calculations of phase dependence for this medium using results presented in Sec. IV and V. 
 It can be seen from Fig. \ref{nonlinearphi} that the self-Kerr nonlinear phase changes are observed even at relatively low light intensity.
  One can observe that the phase gets larger if the optical intensity increases and the phase modulation
   can reach several radians at readily available conditions ($P = 120$ mW, crystal length $L=$100 $\mu$m). As in the case of absorption, the medium exhibits bleaching; the refraction index and corresponding phase shift is decreasing with power thorough the whole spectrum.
The self-Kerr nonlinear optical properties of the system can be controlled by changing the crystal size.
Due to the absorption, there is an interplay between transmissivity and field intensity.
This is shown on the Fig. \ref{nonlinearphi2}. For any desired phase change, there
is a range of parameters for which such a value can be obtained. One can see that while the
 maximum value of $6$ rad is acquired only for very low, practically negligible transmission, the phase
 change of  $\pi$ can be reached with $10\%$ transmission and $P \sim 100$ mW.


\begin{table}[ht!]
\caption{\small Band parameter values for Cu$_2$O, masses in free
electron mass $m_0$.}
\begin{center}
\begin{tabular}{p{.2\linewidth} p{.2\linewidth} p{.2\linewidth} p{.2\linewidth} p{.2\linewidth}}
\hline
Parameter & Value &Unit&Reference\\
\hline $E_g$ & 2172.08& meV& \cite{Kazimierczuk}\\
$R^*$&87.78& meV &\cite{FK}\\
$\Delta_{LT}$&$1.25\times 10^{-3}$&{meV}& \cite{Stolz}\\
$m_e$ & 0.99& $m_0$&\cite{Naka}\\
$m_h$ &0.58&  $m_0$&\cite{Naka}\\
$\mu$ & 0.363 &$m_0$&\\
$\mu'$&-2.33&$m_0$&\\
$M_{tot}$&1.56& $m_0$&\\
$a^*$&1.1& nm&\cite{FK}\\
$r_0$&0.22& nm&\cite{Zielinska.PRB}\\
$\epsilon_b$&7.5 &&\cite{Kazimierczuk}\\
$T_1$&500&ns&\\ \hline
\end{tabular} \label{parametervalues}\end{center}
\end{table}
\section{Conclusions}\label{conclusions} We have developed a simple
mathematical procedure to calculate the nonlinear optical
functions of a semiconductor crystal with Rydberg excitons.
The experiments with Cu$_2$O have made it
possible to observe the nonlinear absorption and dependence of
oscillator strength on laser power, and our theoretical results
show a good agreement with experimental data,
facilitating the calculation of absorption
spectra for any number of excitonic states in a wide range of
conditions (laser power, temperature). Taking
advantage of optical functions in nonlinear regime we have studied
how SPM, which is a measurable evidence of self-Kerr interaction,
can be controlled by laser intensity. Obtained results show that,
depending on the length of the crystal, it is possible to reach a
phase shift of $\pi$ for selected excitonic states even for
relatively low light intensities. We conclude that the Real
Density Matrix Approach is well suited for describing the linear
and nonlinear properties of various types of excitons.
\section*{Acknowledgments}
Support from National Science Centre, Poland (project OPUS, CIREL
2017/25/B/ST3/00817) is greatly acknowledged.
 \end{document}